\documentclass[twocolumn,twoside]{article}
\makeatletter\if@twocolumn\PassOptionsToPackage{switch}{lineno}\else\fi\makeatother

\usepackage{amsfonts,amssymb,amsbsy,latexsym,amsmath,tabulary,graphicx,times,xcolor}

\usepackage[T1]{fontenc}
\usepackage{float}
\usepackage{wrapfig}

\usepackage{url,multirow,morefloats,floatflt,cancel,tfrupee}
\makeatletter
\AtBeginDocument{\@ifpackageloaded{textcomp}{}{\usepackage{textcomp}}}
\makeatother
\usepackage{colortbl}
\usepackage{pifont}
\usepackage[nointegrals]{wasysym}
\makeatletter

\def\mcWidth#1{\csname TY@F#1\endcsname+\tabcolsep}

\def\cAlignHack{\rightskip\@flushglue\leftskip\@flushglue\parindent\z@\parfillskip\z@skip}
\def\rAlignHack{\rightskip\z@skip\leftskip\@flushglue \parindent\z@\parfillskip\z@skip}

\@ifundefined{etal}{}{}

\usepackage{ifxetex}
\ifxetex\else\if@twocolumn\@ifpackageloaded{stfloats}{}{\usepackage{dblfloatfix}}\fi\fi

\AtBeginDocument{
\expandafter\ifx\csname eqalign\endcsname\relax
\def\eqalign#1{\null\vcenter{\def\\{\cr}\openup\jot\m@th
  \ialign{\strut$\displaystyle{##}$\hfil&$\displaystyle{{}##}$\hfil
      \crcr#1\crcr}}\,}
\fi
}

\AtBeginDocument{%
  \@ifpackageloaded{endfloat}%
   {\renewcommand\efloat@iwrite[1]{\immediate\expandafter\protected@write\csname efloat@post#1\endcsname{}}}{\newif\ifefloat@tables}%
}%

\def\BreakURLText#1{\@tfor\brk@tempa:=#1\do{\brk@tempa\hskip0pt}}
\let\lt=<
\let\gt=>
\def\processVert{\ifmmode|\else\textbar\fi}

\@ifundefined{subparagraph}{
\def\subparagraph{\@startsection{paragraph}{5}{2\parindent}{0ex plus 0.1ex minus 0.1ex}%
{0ex}{\normalfont\small\itshape}}%
}{}

\newcommand\role[1]{\unskip}
\newcommand\aucollab[1]{\unskip}
  
\@ifundefined{tsGraphicsScaleX}{\gdef\tsGraphicsScaleX{1}}{}
\@ifundefined{tsGraphicsScaleY}{\gdef\tsGraphicsScaleY{.9}}{}
\def\checkGraphicsWidth{\ifdim\Gin@nat@width>\linewidth
	\tsGraphicsScaleX\linewidth\else\Gin@nat@width\fi}

\def\checkGraphicsHeight{\ifdim\Gin@nat@height>.9\textheight
	\tsGraphicsScaleY\textheight\else\Gin@nat@height\fi}

\def\fixFloatSize#1{}

\DeclareMathAlphabet{\mathpzc}{OT1}{pzc}{m}{it}

\def\URL#1#2{\@ifundefined{href}{#2}{\href{#1}{#2}}}

\def\UrlOrds{\do\*\do\-\do\~\do\'\do\"\do\-}%
\g@addto@macro{\UrlBreaks}{\UrlOrds}

\edef\fntEncoding{\f@encoding}

\makeatother

\newif\ifmultipleabstract\multipleabstractfalse%
\usepackage[paperheight=11in,paperwidth=8.5in,margin=2cm,headsep=.5cm,top=2.5cm,bottom=2.5cm]{geometry}
\usepackage[hang,flushmargin]{footmisc}

\usepackage{abstract}
\renewenvironment{onecolabstract}
{\vspace*{-1pc}\trivlist\item[]\leftskip15pt\rightskip15pt\hrulefill\par\vskip4pt\noindent\textbf{\abstractname:}\mbox{\null}}{\vspace{-6pt}\par\noindent\hrulefill\endtrivlist}

\date{}

\usepackage{caption}
\DeclareCaptionLabelFormat{figlabel}{Fig. #2}

\makeatletter\def\doveIndent{1pt}
\let\@journalTitle\@empty%
\def\journalTitle#1{\gdef\@journalTitle{#1}}

\def\author#1{\gdef\@author{\hskip-\dimexpr(\tabcolsep)\hskip\doveIndent\parbox{\dimexpr\textwidth-\doveIndent}{\centering#1}}}

\def\title#1{%
  \gdef\@title{%
  \vspace{-52pt}%
  \ifx\@journalTitle\@empty\else%
    {\raggedright\fontsize{10}{12}\selectfont\@journalTitle\ 12(12):\ 3090-3097,\ 2017\\ISSN:\ \@issn\\
    [-11pt]\rule{\textwidth}{.2pt}\\}%
  \fi%
  \bfseries\ifx\@articleType\@empty\else\@articleType\\\fi#1%
  }
}

\let\@articleType\@empty \def\articletype#1{\gdef\@articleType{{\normalfont\itshape#1}}}
\let\@issn\@empty \def\issn#1{\gdef\@issn{#1}}
\let\@corresp\@empty
\def\corresp#1{%
  \gdef\@corresp{#1}
}

\usepackage{fancyhdr}
\AtBeginDocument{\usepackage{lastpage}}

\fancypagestyle{plain}{\fancyhf{}\fancyfoot[L]{\ifx\@corresp\@empty\else\@corresp\\\fi\hfill\thepage\hfill}}
\fancypagestyle{headings}{\fancyhf{}\fancyhead[C]{\itshape\@journalTitle\ 12(12):\ 3090-3097,\ 2017}\fancyfoot[C]{\thepage}}

\def\NormalBaseline{\def\baselinestretch{1.1}}

\usepackage{textcase}
\usepackage[explicit]{titlesec}
\titleformat{\section}[hang]{\NormalBaseline\filcenter\large\boldmath\bfseries\small}
{\thesection}
{10pt}
{\noindent\MakeTextUppercase{#1}}
[]
\titleformat{\subsection}[runin]{\NormalBaseline\boldmath\bfseries}
{\thesubsection}
{10pt}
{#1}
[:]
\titleformat{\subsubsection}[runin]{\NormalBaseline}
{\upshape\thesubsubsection}
{10pt}
{#1}
[:]
\titleformat{\paragraph}[runin]{\NormalBaseline}
{\theparagraph}
{10pt}
{#1}
[:]
\titleformat{\subparagraph}[runin]{\NormalBaseline}
{\thesubparagraph}
{10pt}
{#1}
[]

\titlespacing{\section}{0pt}{1.5\baselineskip}{.2\baselineskip}
\titlespacing{\subsection}{0pt}{1.5\baselineskip}{.2\baselineskip}
\titlespacing{\subsubsection}{0pt}{1.5\baselineskip}{.2\baselineskip}
\titlespacing{\paragraph}{0pt}{.5\baselineskip}{10pt}
\titlespacing{\subparagraph}{0pt}{.5\baselineskip}{10pt}

\makeatother

\usepackage[authoryear]{natbib}

\begin{document}

\title{Pothole Detection and Analysis System (PoDAS) for Real Time Data Using Sensor Networks}
\author{\textsuperscript{1}Jinesh Mehta,
            \textsuperscript{1}Vinayak Mathur,
            \textsuperscript{1}Dhruv Agarwal,
            \textsuperscript{1}Atish Sharma and 
            \textsuperscript{1}Krishna Prakasha~\\
\textsuperscript{1}{Department of Information and Communication Technology\unskip, Manipal Institute of Technology, Manipal University\unskip, Manipal\unskip, Karnataka\unskip, India}}\corresp{\textbf{Corresponding Author:} Jinesh Mehta, Department of Information and Communication Technology\unskip, Manipal Institute of Technology, Manipal University\unskip, Manipal\unskip, Karnataka\unskip, India}
\def\RunningHead{{Pothole Detection and Analysis System (PoDAS) for Real Time Data Using Sensor Networks}}\journalTitle{Journal of Engineering and Applied Sciences}\def\RunningAuthor{Mehta \MakeLowercase{\textit{et al.}} }\issn{1816-949X}

\twocolumn[ \maketitle {\begin{onecolabstract}
Potholes are a major nuisance on the city roads leading to several problems and losses in productivity. Local authorities have cited a lack of geographic localization of these potholes as one of the rate-limiting factors for repairs. This study proposes a novel low-cost wireless sensor-based end-to-end system called PoDAS (Pothole Detection and Analysis System) which can be deployed across major cities. We discuss multiple implementation models that can be varied based on the needs of individual cities. Our system uses cross-validation through multiple sensors to achieve higher efficiency than some of the previous models that have been proposed. We also present the results from extensive testing carried out in different environments to ascertain both the efficacy and the efficiency of the proposed system.

\def\keywordstitle{Keywords}

\end{onecolabstract}}]\saythanks 
    
\section*{Introduction}
Technology is best used when it supplements humanity's pursuit of excellence. Across the world, and especially in developing countries, potholes on city roads are a common sight. These potholes not only cost time to the commuters but the loss in productivity translates to a loss of millions of dollars to the global economy. Across cities, local municipalities and governing authorities have reported a lack of information regarding the location and size of these potholes as the single biggest hindrance in their efforts to repair the roads. These problems are augmented during the rainy season when potholes tend to surface sooner. The researchers aim to solve this problem of localization of potholes using a unique combination of wireless sensors, a continuous online data collection technique, and real-time processing of the data rendering an easy-to-use interface for the local authorities. 

We propose a unique solution called Pothole Detection and Analysis System (PoDAS) which can be deployed in large cities in multiple fashions. We discuss two possible application methods in this study. The first application method is to enlist the assistance of radio cabs and mobile application-based taxi aggregators (like Meru and Uber, respectively). The sensor module, which is discussed in greater detail in the successive sections of the study, is attached to the lower end of these taxis which practically traverse the length and breadth of the city multiple times a day. The sensor module takes advantage of the Global Positioning System (GPS) available in the taxi driver's mobile phone. One major hurdle observed in practical implementations is the lack of GPS coordinates to label the captured data. Since these taxi drivers are required to have their GPS modules switched on if they ply on the roads, the sensor system can latch onto that for data labeling. 

Another application model can be used where the local authorities also control the public transportation of a city. In such cases, these modules can be attached to the lower surface of city buses. As the buses ply in the city, the sensor system can gather data and push it to the cloud as soon as it enters an internet zone which can be the Wireless Fidelity (Wi-Fi) at traffic lights in certain cities and the bus depots in others. This caching approach to uploading data on the cloud is explained in detail in the methodology study. 

PoDAS does not use any excessive cost sensors like the Microsoft Kinect sensor which significantly reduces the cost of the sensor modules when compared to other solutions. Our experimentation shows that as we use multiple sensors to collaborate during the data gathering phase and use them to validate the data gathered by individual components, our method achieves significantly better accuracy results. We also provide a very easy interface to the government authorities to understand and locate multiple potholes which does not require any technical knowledge to operate. This makes the proposed system particularly attractive for deployment in developing countries where the government has not necessarily caught up with technological advancements.

 Some of the related work done with respect to pothole detection and road surface monitoring with the help of sensor networks are now discussed. Existing methods for pothole detection are very diverse. Following are the different approaches attempted to solve this problem.

 A system called Pothole Patrol (P2) was introduced by Jakob Eriksson et al. which captures the data from mobility, vibration, and GPS sensors of the participating vehicles and processes that data to assess the road conditions. Potholes and other severe road surface anomalies are identified by applying a simple machine learning approach using X and Z~axis acceleration along with the vehicle's velocity data as input on accelerometer data. The proposed model provides a mobile sensing system which helps in detecting road irregularity using Android OS-based smartphones. Microphone and accelerometer sensors are used to detect potholes and true positive rates obtained are 90\%. The researchers use a tri-axial accelerometer for obtaining the readings of acceleration while riding a motorcycle. Further, the data is processed using various machine learning techniques with respect to several aspects. Supervised and unsupervised methods are used to identify the road quality and road anomalies, and Support Vector Machine (SVM) is used to identify the respective position of the detected anomalies. With the help of these methods and using Kendall tau rank correlation coefficient, an accuracy of 78.5\% is achieved when the comparison is done with the data obtained by actual human observation with the data obtained after applying the above machine learning methods. 

Traffic Sense and Nericell are the systems developed by Prashanth Mohan et al. which use various sensing components like accelerometer, microphone, Global System for Mobile (GSM) radio, and GPS sensor. Identification of potholes, bumps, braking, and honking is done with the help of the above sensors. The researchers build a network called BusNet which considers a public transport system-based sensor network to monitor road surface conditions and environmental pollution. Road Condition Monitoring App developed by Avik Ghose et al. is a monitoring and alert system for the road conditions of a city. This app uses in-vehicle smartphones as connected sensors which are connected to the Internet with the help of the Internet of Things (IoT). This approach is energy-efficient as it reduces the data communication between the phone and back-end system and provides accurate road conditions as many users are used to authenticate them. 

Microsoft Kinect and a high-speed Universal Serial Bus (USB) camera are used as sensing devices. Mednis et al. propose a methodology for the detection of potholes with the help of mobile vehicles rigged with shelf microphones and global positioning devices connected to an onboard computer. Whereas metrological and visualization properties of a pothole are discussed by Moazzam et al. where a Kinect sensor is used which helps in collecting pavement depth images of roads to analyze the area of potholes with respect to depth. The estimated volume of potholes is measured using the trapezoidal rule on area-depth curves through pavement image analysis. 

CarMote is a customized embedded device dedicated to monitoring road surfaces using microphone and accelerometer sensors. A novel pothole detection system was proposed by Rode et al. which helps the driver in avoiding potholes on the roads with the help of prior warnings. The architectural design further proposes a low response time, minimal maintenance, and deployment cost solution to this problem.

Most people who drive on roadways are the ones who suffer from the presence of potholes. There are various problems these potholes affect them in many ways: first is the maintenance required due to a vehicle passing over these potholes. If potholes are detected and recorded at regular intervals, then the damage incurred due to these potholes can be reduced. If the municipal cooperation in charge of repairing potholes was made aware of pothole locations in a timely manner, then they could be more quickly filled, resulting in less vehicle driving over them and causing further damage to the road and their vehicles. PoDAS aims to do this and provides the following aspects:

\begin{itemize}
\item We use an ultrasonic sensor which provides high accuracy to detect potholes.
\item Real-time data sent to the cloud where data is analyzed.
\item Visual interpretation of the potholes and uneven roads on Google Maps.
\end{itemize}

\section*{Materials and Methods}
\subsection*{Proposed Algorithm}
The complete model proposed for detecting potholes in this study is shown in Fig. 1. Sensors given in the figure are a combined circuit with all the four components: ultrasonic sensor, Arduino Uno, GPS module, and accelerometer. It interacts with the local server which communicates to the back-end server with the help of the internet which acts as a transmission medium. Data computation is done on the back-end server and processed data is passed to the front-end application. The user interacts with this application and can get a visual idea of the quality of the road. 

\begin{figure}[h]
\includegraphics[width=\linewidth]{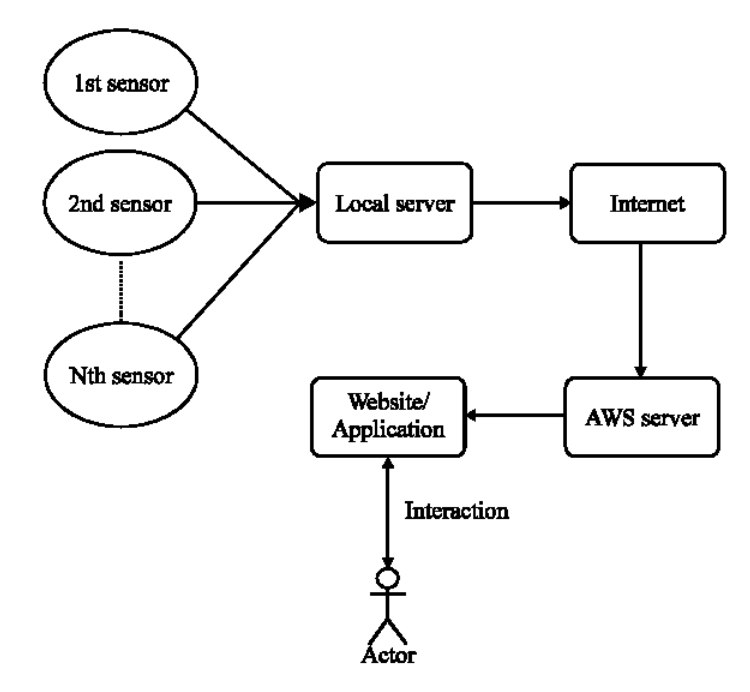}
\caption{Schematic representation of the proposed Pothole Detection and Analysis System (PoDAS).}
\label{fig:schematic}
\end{figure}

The entire process is divided into different segments which are narrowed down to two segments:

\begin{itemize}
\setlength\itemsep{0em}
\item Data acquisition
\item Data processing
\end{itemize}

Data acquisition is the use of sensors to sense the environment and report the collected data values back to a host. Whereas data processing is the host consolidating the data and deriving meaning out of it. Later part of this section explains both segments in detail. In order to implement the proposed solution, the following sensors are used each working independently of the other and providing respective sensed values:

\begin{itemize}
\setlength\itemsep{0em}
\item Ultrasonic sensor: HC-SR04 Distance Measuring Transducer sensor
\item Arduino Uno R3 Compatible with USB Cable
\item GPS module NEO6MV2 NEO-6M
\item GY-521 Mpu6050 Module + Accelerometer for Arduino
\end{itemize}

\subsection*{Data Acquisition}
The ultrasonic sensor is the first and primary source of pothole detection. The sonic waves are generated by the transducer which strikes with an object (in this case the road), gets reflected, and comes back to the transducer. After having transmitted the sonic waves, the ultrasonic sensor will change to receiving mode. The amount of time taken between transmitting and receiving is proportional to the distance of the object from the sensor. Thus, we get the distance value which is converted into inches. A particular threshold value is calculated by taking some constant readings of a new road or well-maintained road. The threshold value computed is stored into the dataset.

The secondary source of detection of potholes is the accelerometer which helps in case the ultrasonic sensor fails to detect an approaching pothole. As soon as the vehicle encounters a pothole, it would inadvertently get displaced across its Z-axis. The accelerometer continuously records these displacements. Similar to the ultrasonic sensor, a particular threshold value is calculated by taking some constant readings of a new road or well-maintained road and that value is stored in the dataset. Once the sensor senses the data for a particular point, the corresponding accelerometer and ultrasonic sensor values of that point are pinned using geographical coordinates of the location and stored in the dataset.

At regular intervals, this information is then sent to the back-end server where all the data is processed and further analysis of the road can be done. If a sensor is not connected to the local server, it stores its value in the cache memory of the sensor and sends it whenever it reconnects to the internet. Similarly, if the local server is not connected to the internet, then the data is buffered by queuing it and sends it to the back-end server as soon as it gets connected to the internet. The circuit diagram of the sensors used to collect the data is shown in Fig. 2. Arduino Uno is connected to three sensors. First is the ultrasonic sensor for keeping track of the ground clearance for a vehicle, next is the accelerometer for keeping track of the movement and orientation of the vehicle, and the last part is the GPS module for saving the location for detected potholes.

\subsection*{Proposed Algorithm for PoDAS}
Now we propose the new algorithm to detect the potholes with higher accuracy which is as given below:

\includegraphics[width=\linewidth]{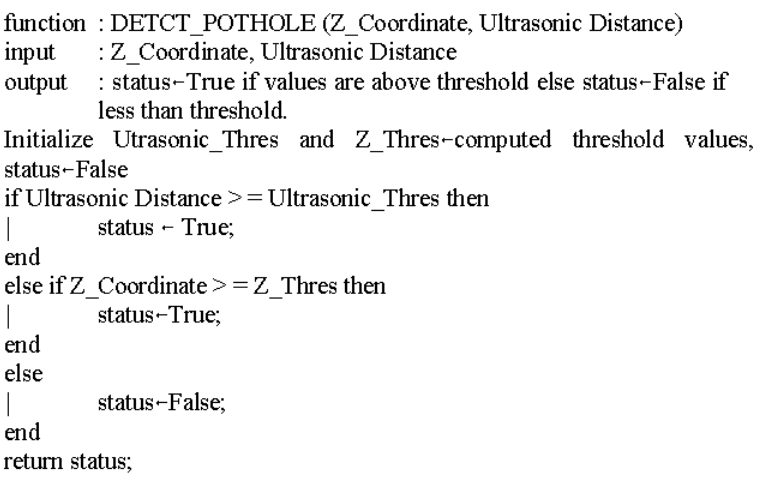}

\begin{figure*}[!h]
    \centering
    \includegraphics[width=\textwidth]{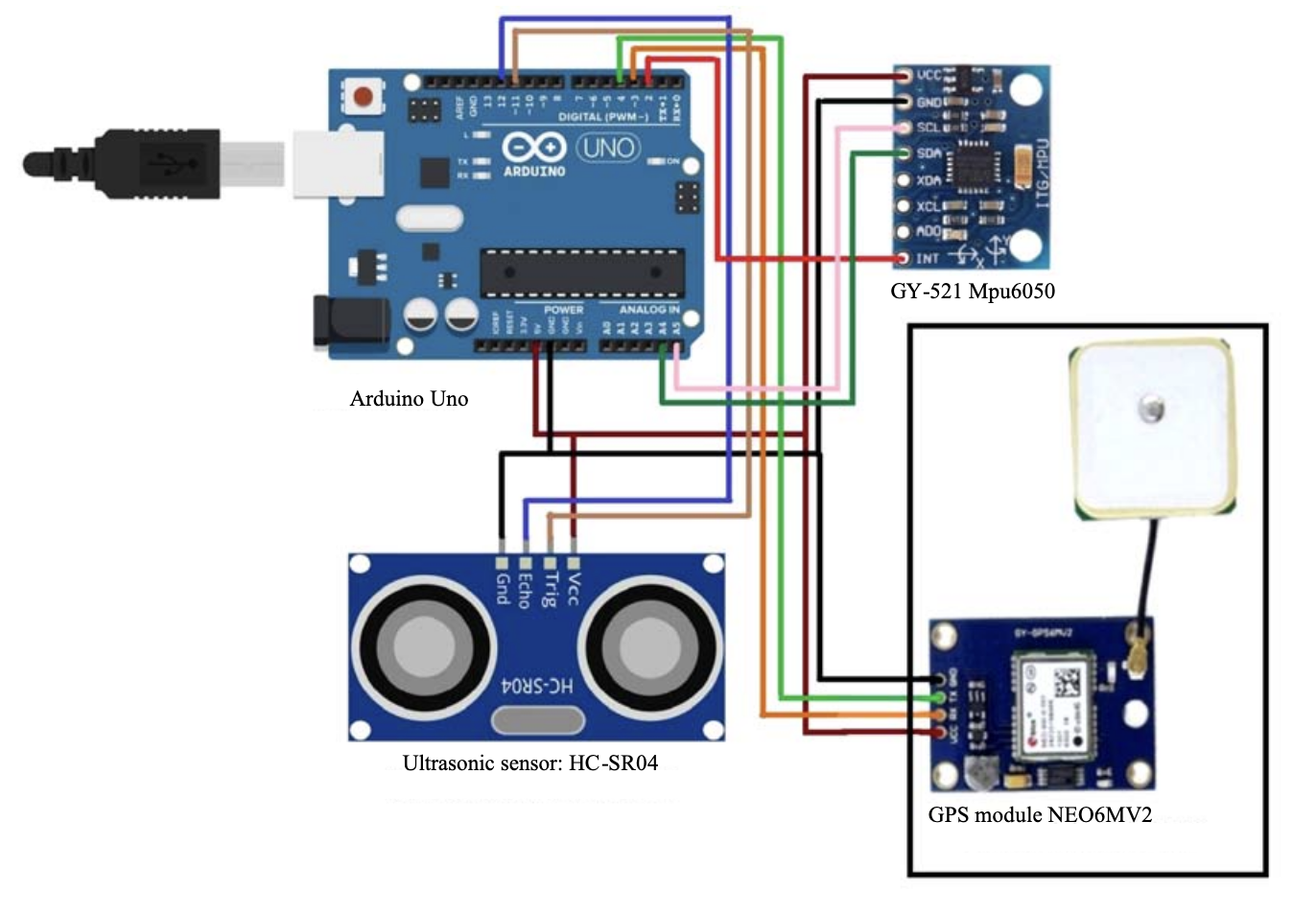}
    \caption{Circuit diagram for data collection using Arduino Uno, ultrasonic sensor, GPS module and GY-521 Mpu6050 (accelerometer).}
    \label{fig:circuit}
\end{figure*}

\subsection*{Data Processing}
The system follows a star topology architecture. Data is provided by the sensors to the server which in turn notifies its clients. The accelerometer threshold values for the Z-axis and the ultrasonic threshold value are retrieved from the dataset. Now the values of the threshold for ultrasonic as well as the accelerometer are compared to readings of all the points in the dataset and hence they are classified as whether there is a pothole on that particular point or not. Also, the threshold values are stored after the first computation and thus next time whenever the data comes, it doesn't need to wait for retrieving these values again from the dataset.

\subsection*{Information Retrieval and Processing}
The sensors send the detected values to the local server which communicates with the back-end server via the internet. Data is marshaled and transmitted as JSON (JavaScript Object Notation). This reduces the size of the data to be transmitted while maintaining the intended structure of the data. The back-end determines if the detected readings are that of a pothole or not. Accordingly, data is sent to its clients. Clients can also query the server as the communication is bi-directional.

\subsection*{Using Cloud Facilities}
As the server is hosted on the cloud, classification of detection of potholes is done over there. All the computation and the proposed algorithm are implemented over here. It is the server's duty to detect if a pothole has been found or not by running the algorithm on the given dataset. All the results are stored in the cloud itself so that it can be passed on to the client.

\subsection*{Reporting Information}
Information is reported back to the client as JSON. This information can then be used to represent pothole locations through charts, graphs, or interactive live maps. This helps the user to retrieve the processed information and take the desirable actions as per the requirement.

\begin{figure*}[b]
    \centering
    \includegraphics[width=\textwidth]{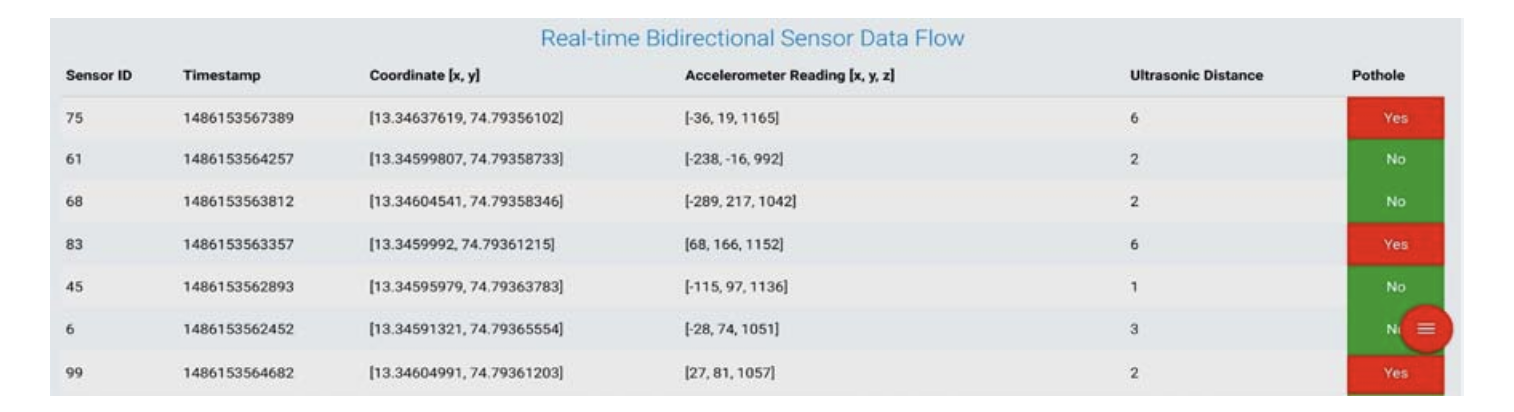}
    \caption{Real-time data obtained from the sensors which is processed at the back-end server and output of whether a pothole is there at a particular coordinate is displayed.}
    \label{fig:realtime}
\end{figure*}

\begin{figure*}[b]
    \centering
    \includegraphics[width=\textwidth]{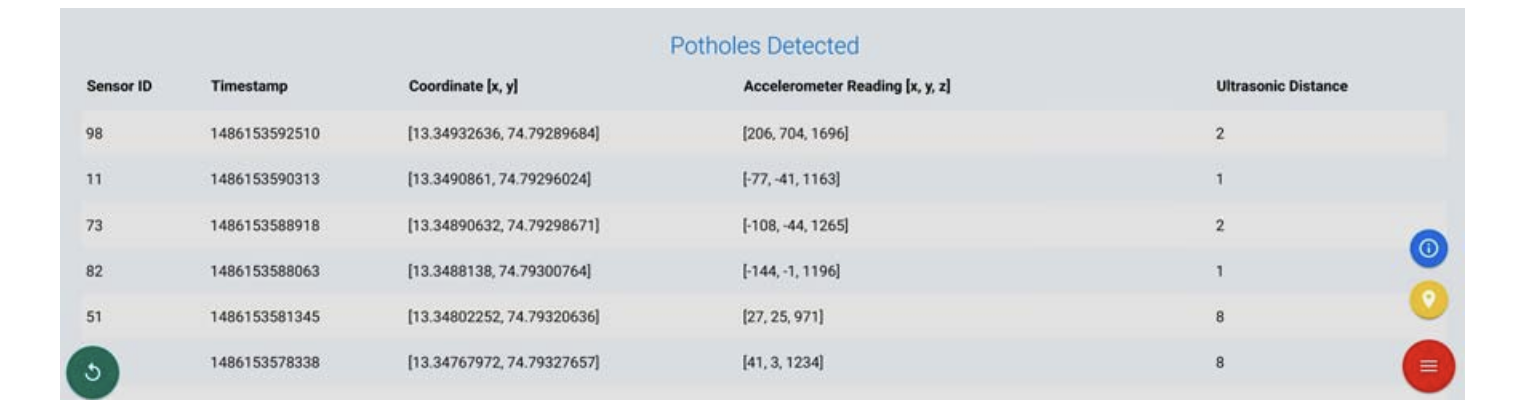}
    \caption{List view of detected potholes which helps to keep a track of all the detected potholes at a single place.}
    \label{fig:listview}
\end{figure*}

\subsection*{Applications}
To apply PoDAS in real-time, a feedback system can be implemented. This ensures that information with the access point is correctly and timely updated. Here are a few modes of applications we present for the deployment of sensors:

\subsection*{Rental cabs}
In this case, the sensor is deployed in every vehicle. As the driver keeps the GPS and internet on the whole time, sensors can be connected to the internet by creating a Wireless Internet hotspot from the driver's phone. Hence, real-time data keeps on flowing into the server at any given time.

\subsection*{Public transport}
This approach proposes the deployment of sensors in public transports like buses. Buses traveling the entire day can store the data into the cache memory of the sensor and as soon as it reaches the bus depot, it gets connected to the internet and all the data is transmitted to the server.

\subsection*{Pilot vehicle}
Vehicles which are designed for purposes can be considered under this category. It moves through the roads at periods and aggregates data about potholes to the system. Internet connection must be provided as and when used.

\section*{Result and Discussion}

\subsection{Experiments and Discussion}
\begin{figure*}[t]
    \centering
    \includegraphics[width=\textwidth]{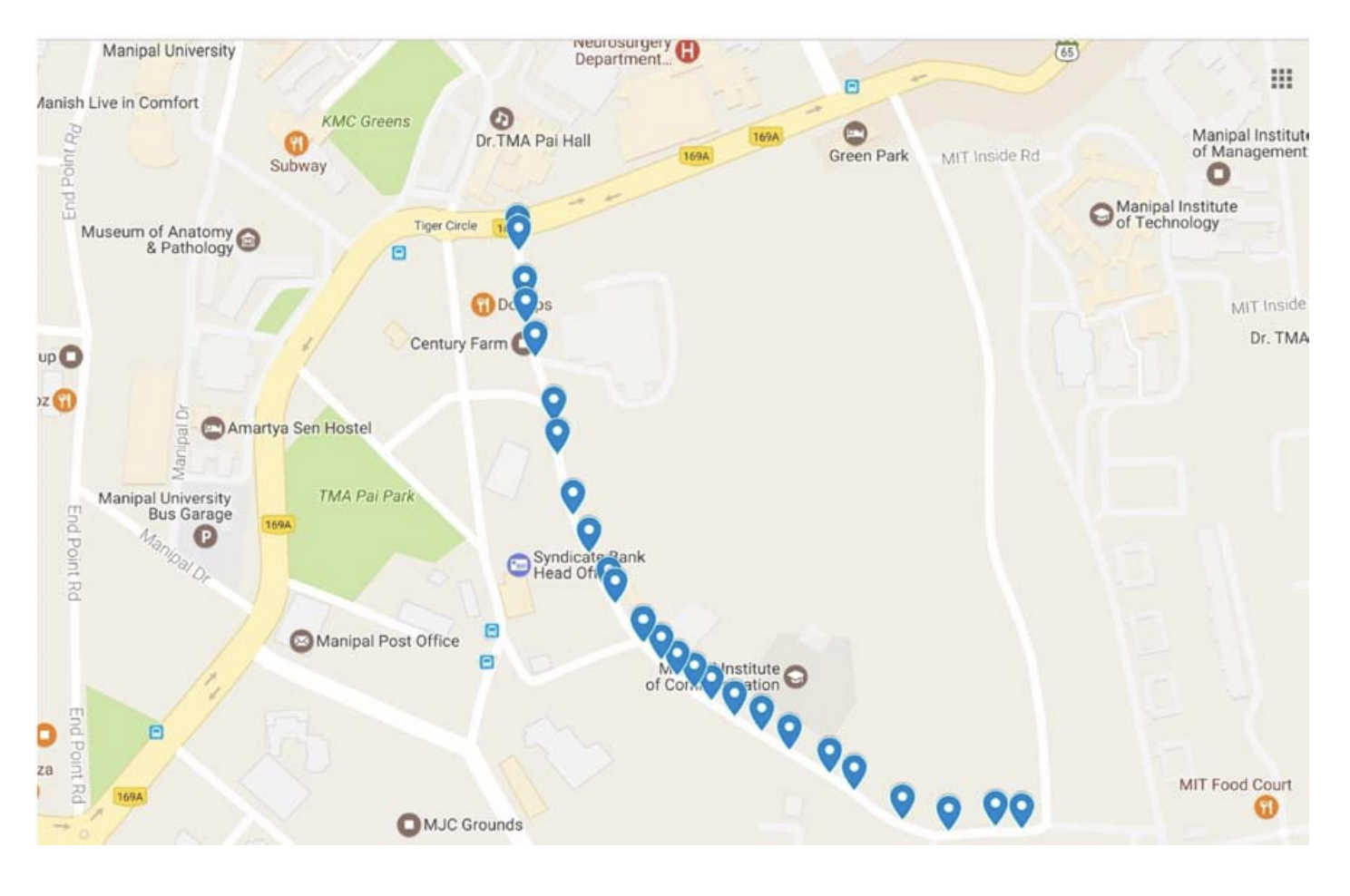}
    \caption{Pinned location for all the detected potholes using PoDAS on Google Maps (Li and Zhijian, 2010).}
    \label{fig:googlemaps}
\end{figure*}

\begin{figure}
    \centering
    \includegraphics[width=\linewidth]{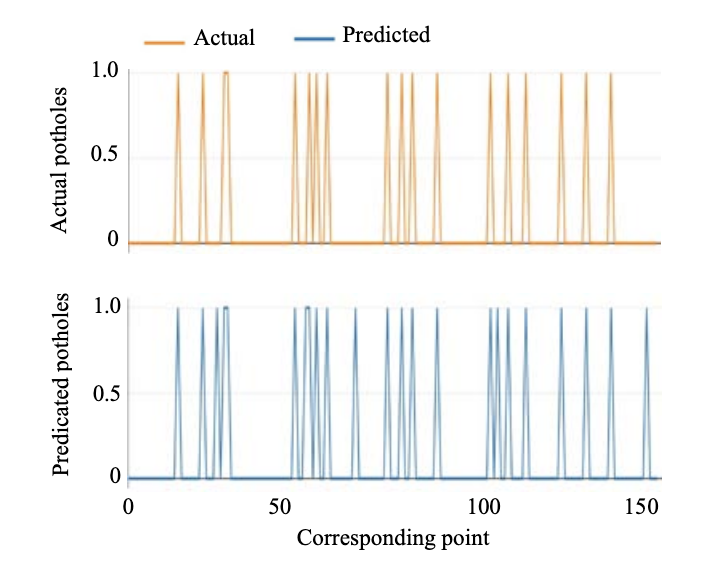}
    \caption{Comparing the predicted potholes and actual potholes. PoDAS is successful in detecting 80\% of the potholes present on the roads during our experimentation.}
    \label{fig:comparison}
\end{figure}

To check the accuracy and efficiency of PoDAS we conducted an experiment. All the sensors were attached as a whole to the bottom of a vehicle and the power was supplied to the Arduino Uno with the help of USB. Readings were saved on a laptop which is connected through USB. The laptop acts as a local server and as soon as it gets connected to the internet the data is transmitted to the Amazon Web Service back-end server where all the processing is done. For the sake of convenience, we collected the data for potholes along the road which was around 1 km long and around 150 points were taken into consideration while traveling on this road. In order to compute the threshold values for the ultrasonic sensor and accelerometer, another road which is high maintained and about 500 m long was taken into consideration. Constant threshold values for ultrasonic sensors and z-coordinate for the accelerometer are obtained to be 6 inches and 1150 m/sec². It was found that a standard default distance of between 6-10 inches indicates the need for maintenance of the road. The reported distance of <10 inches indicates the presence of a large pothole that needs immediate attention by the authorities.

This information obtained as shown in Fig. 3 is viewed onto the system to issue the warning of an upcoming pothole or to initialize road maintenance. This makes it easy for keeping track of the quality of roads especially after heavy rains as the roads get damaged more during that time. To get a list of all the potholes in a particular place, a different web page was created which is shown in Fig. 4. With this information obtained from the detected pothole webpage, a time series can be obtained which can help the government to determine the budget for road maintenance. For better interpretation of the data, Fig. 5 is used which helps the road maintenance people to locate the approximate location on Google Maps. To obtain this representation, all the recent detected potholes are passed to the Google Maps API which pinpoints all the locations provided to it via GPS coordinates (Fig. 6).

\section*{Conclusion}
We obtained 25 potholes while performing the above experiment and out of 25 potholes, 20 of them are detected by our method. This suggests an accuracy of 80\% which is very high as compared to other methods. Comparison of the predicted potholes and actual potholes can be seen. In conclusion, we aimed to address a problem faced by many local municipal authorities where they do not know the exact geographic locations of potholes on the roads. In seasons where there is a lot of precipitation, the situation is more dynamic and it is much harder for the authorities to keep track of new potholes. In the past, they have relied on citizen initiatives to gain such information. Given the economic ramifications of these potholes, we have come up with a low-cost technology alternative that will help the concerned government authorities. Our proposed PoDAS system can be deployed in multiple ways depending on the city's needs. We use low-cost sensors which make the system economically attractive. We perform extensive experiments as well as real-world testing which show that our system is able to detect potholes of all shapes and sizes with an accuracy of roughly 80\%. In the future, this setup could be bundled into standalone devices for commercialization. Different sensor calibration can be explored for further improving the accuracy obtained.

\section*{References}
\begin{itemize}
\item Deepika, K. and J. Usha, 2016. Investigations and implications on location tracking using RFID with global positioning systems. Proceedings of the 3rd International Conference on Computer and Information Sciences (ICCOINS), August 15-17, 2016, IEEE, Kuala Lumpur, Malaysia, ISBN: 978-1-5090-25503, pp: 242-247.
\item El-Aziz, A.A. and A. Kannan, 2014. JSON encryption. Proceedings of the 2014 International Conference on Computer Communication and Informatics (ICCCI), January 3-5, 2014, IEEE, Coimbatore, India, ISBN: 978-1-4799-2352-6, pp: 1-6.
\item Eriksson, J., L. Girod, B. Hull, R. Newton, S. Madden and H. Balakrishnan, 2008. The pothole patrol: Using a mobile sensor network for road surface monitoring. Proceedings of the 6th International Conference on Mobile Systems, Applications and Services, June 17-20, 2008, Breckenridge, Colorado, pp: 29-39.
\item Ghose, A., P. Biswas, C. Bhaumik, M. Sharma and A. Pal et al., 2012. Road condition monitoring and alert application: Using in-vehicle smartphone as internet-connected sensor. Proceedings of the IEEE International Conference on Pervasive Computing and Communications Workshops (PERCOM Workshops), March 19-23, 2012, IEEE, Lugano, Switzerland, ISBN: 978-1-4673-0905-9, pp: 489-491.
\item Li, H. and L. Zhijian, 2010. The study and implementation of mobile GPS navigation system based on Google Maps. Proceedings of the 2010 International Conference on Computer and Information Application (ICCIA), December 3-5, 2010, IEEE, Tianjin, China, ISBN: 978-1-4244-8597-0, pp: 87-90.
\item Lindqvist, U. and P.G. Neumann, 2017. The future of the internet of things. Commun. ACM., 60: 26-30.
\item Matijevic, M. and V. Cvjetkovic, 2016. Overview of architectures with Arduino boards as building blocks for data acquisition and control systems. Proceedings of the 2016 13th International Conference on Remote Engineering and Virtual Instrumentation (REV), February 24-26, 2016, IEEE, Madrid, Spain, ISBN: 978-1-4673-8245-8, pp: 56-63.
\item Mednis, A., A. Elsts and L. Selavo, 2012. Embedded solution for road condition monitoring using vehicular sensor networks. Proceedings of the 6th International Conference on Application of Information and Communication Technologies (AICT), October 17-19, 2012, IEEE, Tbilisi, Georgia, ISBN: 978-1-4673-1739-9, pp: 1-5.
\item Mednis, A., G. Strazdins, M. Liepins, A., Gordjusins and L. Selavo, 2010. Roadmic: Road surface monitoring using vehicular sensor networks with microphones. Proceedings of the International Conference on Networked Digital Technologies, July 7-9, 2010, Springer, Berlin, Germany, pp: 417-429.
\item Mednis, A., G. Strazdins, R. Zviedris, G. Kanonirs and L. Selavo, 2011. Real-time pothole detection using android smartphones with accelerometers. Proceedings of the 2011 International Conference on Distributed Computing in Sensor Systems and Workshops (DCOSS), June 27-29, 2011, IEEE, Barcelona, Spain, ISBN: 978-1-4577-0512-0, pp: 1-6.
\item Moazzam, I., K. Kamal, S. Mathavan, S. Usman and M. Rahman, 2013. Metrology and visualization of potholes using the Microsoft Kinect sensor. Proceedings of the 16th International IEEE Conference on Intelligent Transportation Systems (ITSC 2013), October 6-9, 2013, IEEE, New York, USA, ISBN: 978-1-4799-2914-6, pp: 1284-1291.
\item Mohan, P., V.N. Padmanabhan and R. Ramjee, 2008. Nericell: Rich monitoring of road and traffic conditions using mobile smartphones. Proceedings of the 6th ACM Conference on Embedded Network Sensor Systems, November 5-7, 2008, ACM, Raleigh, North Carolina, USA, ISBN: 978-1-59593-990-6, pp: 323-336.
\item Mohan, P., V.N. Padmanabhan and R. Ramjee, 2008. Nericell: Using mobile smartphones for rich monitoring of road and traffic conditions. Proceedings of the 6th ACM Conference on Embedded Network Sensor Systems, November 5-7, 2008, ACM, Raleigh, North Carolina, USA, ISBN: 978-1-59593-990-6, pp: 357-358.
\item Nguyen, T.S., T.H. Tran and H. Vu, 2016. Accurate object localization using RFID and Microsoft Kinect sensor. Proceedings of the 2016 8th International Conference on Knowledge and Systems Engineering (KSE), October 6-8, 2016, IEEE, Hanoi, Vietnam, ISBN: 978-1-4673-8929-7, pp: 345-350.
\item Rode, S.S., S. Vijay, P. Goyal, P. Kulkarni and K. Arya, 2009. Pothole detection and warning system: Infrastructure support and system design. Proceedings of the International Conference on Electronic Computer Technology, February 20-22, 2009, IEEE, Macau, China, ISBN: 978-0-7695-3559-3, pp: 286-290.
\item Su, K.C., H.M. Wu, W.L. Chang and Y.H. Chou, 2012. Vehicle-to-vehicle communication system through Wi-Fi network using Android Smartphone. Proceedings of the 2012 International Conference on Connected Vehicles and Expo (ICCVE), December 12-16, 2012, IEEE, Beijing, China, ISBN: 978-1-4673-4705-1, pp: 191-196.
\item Tong, S. and D. Koller, 2001. Support vector machine active learning with applications to text classification. J. Mach. Learn. Res., 2: 45-66.
\item Vittorio, A., V. Rosolino, I. Teresa, C.M. Vittoria and P.G. Vincenzo, 2014. Automated sensing system for monitoring of road surface quality by mobile devices. Procedia Soc. Behav. Sci., 111: 242-251.
\item Zoysa, K.D., C. Keppitiyagama, G.P. Seneviratne and W.W. A.T. Shihan, 2007. A public transport system-based sensor network for road surface condition monitoring. Proceedings of the 2007 Workshop on Networked Systems for Developing Regions, August 27-27, 2007, ACM, Kyoto, Japan, ISBN: 978-1-59593-787-2, pp: 1-9.
\end{itemize}

\bibliographystyle{plainnat}

\end{document}